\documentclass[10pt,conference]{IEEEtran}
\def\BibTeX{{\rm B\kern-.05em{\sc i\kern-.025em b}\kern-.08em
    T\kern-.1667em\lower.7ex\hbox{E}\kern-.125emX}}
\IEEEoverridecommandlockouts 
\usepackage{blkarray}                                      
\usepackage{algpseudocode}                                 
\usepackage{algorithm}
\usepackage{graphicx}                                      
\usepackage{amsmath}
\usepackage{amssymb}
\usepackage{amsfonts}
\usepackage{amsthm}
\usepackage[mathcal]{eucal}
\usepackage{mathrsfs}
\usepackage{booktabs}
\usepackage{enumerate}
\usepackage{multirow}
\usepackage{color}
\usepackage{cite}                                          
\usepackage{comment}                                       
\usepackage{soul}                                          
\soulregister\cite7
\soulregister\ref7
\soulregister\pageref7
\usepackage{etoolbox}                                      
\usepackage{url}
\usepackage{nth}                                           
\usepackage{bm}                                            
\usepackage{courier}
\usepackage{balance}
\usepackage{threeparttable}
\usepackage{xcolor,colortbl}
\usepackage{footnote}
\usepackage{listings}
\usepackage{setspace}                                      
\usepackage[inline]{enumitem}

\usepackage{verbatim}
\usepackage[bookmarks=false]{hyperref}
\hypersetup{
    colorlinks = true,
    citecolor  = blue,
    linkcolor  = blue,
    urlcolor   = blue,
}
\usepackage{tikz}
\usetikzlibrary{patterns,snakes}
\usetikzlibrary{positioning,calc,fit,decorations.pathmorphing,shapes.geometric, shapes.gates.logic.US, calc}
\usetikzlibrary{arrows,arrows.meta,decorations.markings,shapes,shapes.arrows}
\usetikzlibrary{decorations,decorations.pathreplacing}
\usetikzlibrary{backgrounds}
\usepackage{filecontents}                                  
\usepackage{pgfplots}
\usepackage{pgfplotstable}
\usepackage{scalefnt}
\pgfplotsset{compat=newest}
\usepackage{caption}
\usepackage{pifont}                                        
\usepackage{cleveref}
\Crefformat{figure}{Fig.~#2#1#3}                           
\Crefname{subfigure}{Fig.}{Figs.}
\Crefname{figure}{Fig.}{Figs.}
\Crefformat{table}{TABLE~#2#1#3}                           
\usepackage[figuresright]{rotating}

\definecolor{CUHKorange}{RGB}{244,106,18} 
\definecolor{CUHKblue}{RGB}{0,111,190}    
\definecolor{CUHKgreen}{RGB}{0,127,128}   
\definecolor{CUHKred}{RGB}{228,46,36}     
\definecolor{CUHKyellow}{RGB}{198,148,34} 
\definecolor{CUHKdark}{RGB}{114,44,114}   
\definecolor{CUHKmiddle}{RGB}{144,44,144} 
\definecolor{CUHKlight}{RGB}{167,44,167} 
\definecolor{CUHKpurple}{RGB}{117,15,109}
\definecolor{CUHKgold}{RGB}{221,163,0}
\definecolor{CUHKribbon}{RGB}{244,223,176}
\definecolor{CUHKblack}{RGB}{34,24,21}

\renewcommand{\bf}[1]{\textbf{#1}}
\renewcommand{\tt}[1]{\texttt{#1}}

\newcommand{\minisection}[1]{\vspace{.06in}\noindent{\textbf{#1}}.}

\usepackage{tcolorbox}
\tcbuselibrary{skins,breakable}
    {\endtcolorbox}
    {\endtcolorbox}

\iftrue
\newcommand{\subparagraph}{}
\usepackage{titlesec}
\titlespacing*{\section}{0pt}{1.8ex plus .2ex minus .2ex}{0.4ex plus .2ex}
\titlespacing*{\subsection}{0pt}{1.0ex plus .2ex minus .2ex}{0.2ex plus .2ex}
\fi

\newtheorem{myproblem}{\textbf{Problem}}

\crefname{mytheorem}{Theorem}{Theorems}
\crefname{mylemma}{Lemma}{Lemmas}
\crefname{myclaim}{Claim}{Claims}
\crefname{myproperty}{Property}{Properties}
\crefname{mycorollary}{Corollary}{Corollaries}

\algrenewcommand\textproc{\texttt}

\makeatletter
\let\OldStatex\Statex
\renewcommand{\Statex}[1][3]{%
  \setlength\@tempdima{\algorithmicindent}%
  \OldStatex\hskip\dimexpr#1\@tempdima\relax
}
\makeatother

\RequirePackage[normalem]{ulem} 
\RequirePackage{color}\definecolor{RED}{rgb}{1,0,0}\definecolor{BLUE}{rgb}{0,0,1} 

\usepackage{times}
\usepackage{listings}
\usepackage{balance}

\definecolor{myorange}{RGB}{238,97,42}  %
\definecolor{myblue}{RGB}{178,179,249}  
\definecolor{mygrey}{RGB}{166,166,166}  %
\definecolor{mygreen}{RGB}{180,210,36}  
\definecolor{myred}{RGB}{238,100,100}   %
\definecolor{myyellow}{RGB}{198,148,34} 
\definecolor{mydark}{RGB}{114,44,114}   
\definecolor{mymiddle}{RGB}{144,44,144} 
\definecolor{mylight}{RGB}{167,44,167}  
\definecolor{myblue1}{RGB}{137,157,192}  
\definecolor{mygreen1}{RGB}{69,137,148}  
\definecolor{mypurple}{RGB}{149,143,226}
\definecolor{myorange1}{RGB}{246,163,122}

\usepackage{textcomp}
\usepackage{xcolor}
\usepackage{comment}
\usepackage{booktabs}
\usepackage{multirow}
\usepackage{array}
\usepackage{subcaption}

\usepackage[a4paper, total={184mm,239mm}]{geometry}

\begin{document}
\date{}

\title{
    Timing-driven Approximate Logic Synthesis Based on Double-chase Grey Wolf Optimizer    
}

\author{
    Xiangfei Hu$^{1}$, 
    Yuyang Ye$^{2}$,  
    Tinghuan Chen$^{2,3}$, 
    Hao Yan$^{1}$, 
    Bei Yu$^{2}$ \\
    $^1$Southeast University \quad
    $^2$CUHK \quad
    $^3$CUHK-Shenzhen \quad \\

    \thanks{
    	This work is accepted by Design, Automation \& Test in Europe Conference (DATE 2025). The corresponding authors are Yuyang Ye and Hao Yan.
    }
}

\maketitle
\pagestyle{plain}
\begin{abstract}
With the shrinking technology nodes, timing optimization becomes increasingly challenging.
Approximate logic synthesis (ALS) can perform local approximate changes (LACs) on circuits to optimize timing with the cost of slight inaccuracy.
However, existing ALS methods that focus solely on critical path depth reduction (depth-driven methods) or area minimization (area-driven methods) are inefficient in achieving optimal timing improvement.
In this work, we propose an effective timing-driven ALS framework,
where we employ a double-chase grey wolf optimizer to explore and apply LACs, simultaneously bringing excellent critical path shortening and area reduction under error constraints.
Subsequently, it utilizes post-optimization under area constraints to convert area reduction into further timing improvement, thus achieving maximum critical path delay reduction.
According to experiments on open-source circuits with TSMC 28nm technology, compared to the SOTA method, our framework can generate approximate circuits with greater critical path delay reduction under different error and area constraints.
\end{abstract}
\section{Introduction}
\label{sec:Intro}
Timing optimization is crucial in VLSI design.
As the CMOS technology nodes continue to shrink, timing improvements caused by traditional methods, including gate sizing and logic restructure, are 
limited \cite{esmaeilzadeh2011dark, han2013approximate}.
In recent years, error-tolerant applications are becoming increasingly popular.
Consequently, approximate computing \cite{han2013approximate}, which effectively balances accuracy and performance, has garnered great attention.
It can significantly reduce circuit delay, area, and power with the cost of slight computational imprecision.

\begin{figure}[tb!]
    \centering
    \includegraphics[width=1\linewidth]{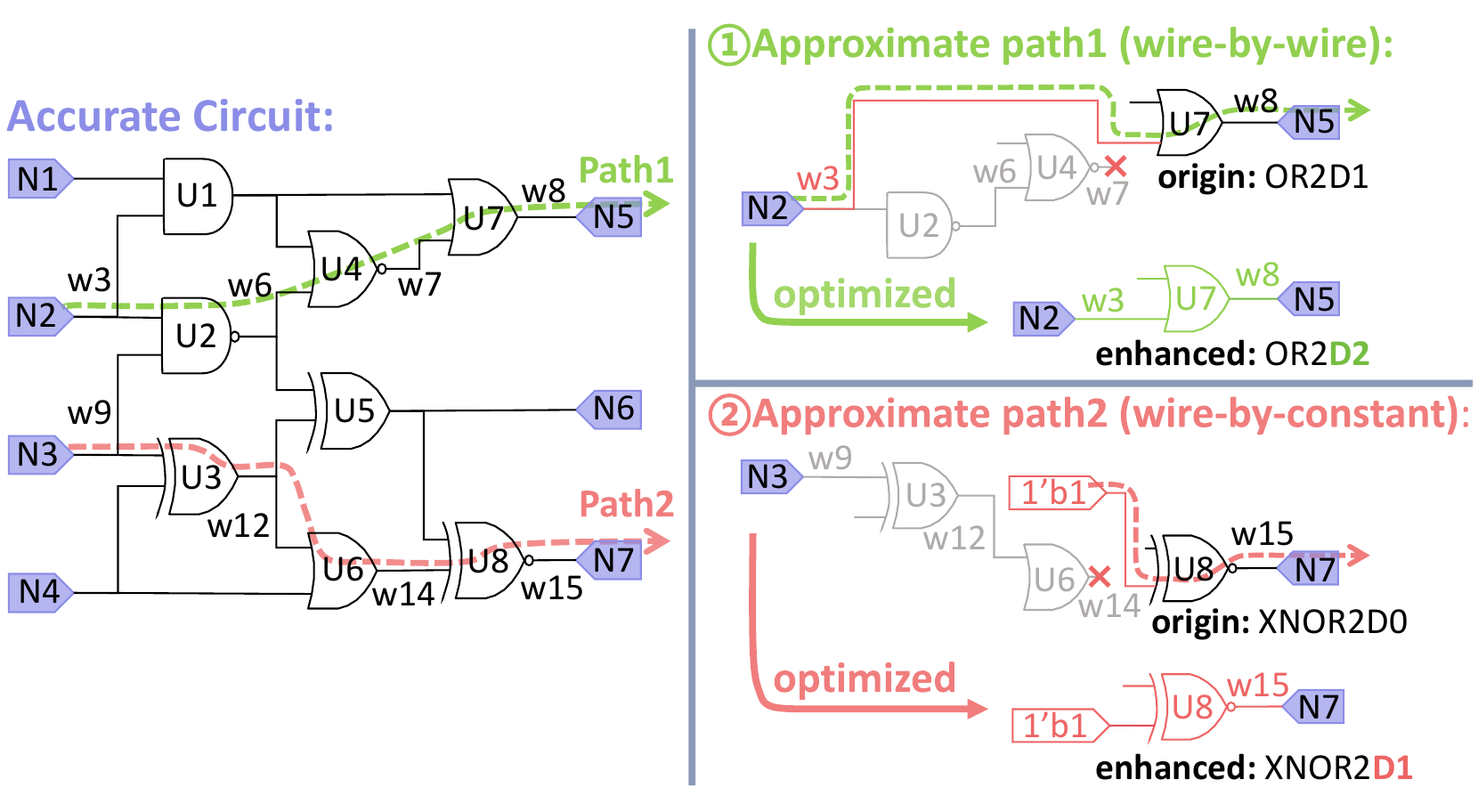}
    \caption{Optimizing circuit by wire-by-wire (substitute a wire with another wire in circuits) and wire-by-constant (substitute a wire with constant logic value `0'$/$`1') LACs. Area reductions are converted into drive strength enhancement of gates.}
    \label{gca}
\vspace{-0.03in}
\end{figure}

Recently, approximate logic synthesis (ALS) has been proposed as an automated approximate computing paradigm.
It can optimize timing under a relaxed error bound by reducing the depth of critical paths and enhancing the drive strength of gates on critical paths \cite{ye2024timing}.
Based on optimization approaches, existing ALS methods can primarily be divided into two categories: (1) depth-driven methods \cite{balaskas2021automated,balaskas2022variability,meng2023hedals} and (2) area-driven methods \cite{meng2020alsrac,meng2022seals,su2022vecbee,lee2023approximate}.
Depth-driven methods perform local approximate changes (LACs) to simplify gates on critical paths, providing direct timing improvement.
As shown in \Cref{gca}, LACs are applied to critical paths 1 and 2.
By omitting certain gates, both paths become shallower and faster with the cost of a slight error.
HEDALS \cite{meng2023hedals} proposes a critical error graph to accelerate critical path depth reduction and strictly control the introduced errors.
Area-driven methods select LACs with the best area reduction potential to minimize circuit area. 
SEALS \cite{meng2022seals} and VECBEE-SASIMI \cite{su2022vecbee} combine fast error estimation with greedy algorithms to iteratively select such LACs, efficiently reducing circuit area.
\Cref{gca} also illustrates that these area reductions can be converted into the enhancement of gate drive strength by post-optimization, leading to further timing improvement.

However, achieving ALS with the greatest potential for timing optimization is challenging for previous methods. 
Specifically, depth-driven methods inadequately reduce area, leading to difficulties in maximizing the drive strength of gates on critical paths.
Area-driven methods simplify many gates on non-critical paths to reduce area, which makes it difficult to obtain the optimal critical path depth.
Therefore, it is necessary for timing-driven ALS to simultaneously optimize both critical path depth and area. 
In this scenario, conventional gradient-based optimizers, including greedy algorithm, genetic algorithm, and traditional grey wolf optimizer (GWO) \cite{mirjalili2016multi} using a single-chase strategy, cannot finely partition the sampled approximate solutions.
Thus, solutions are dispersed in the solution space.
This dispersion causes an excessive number of gradients for further optimization.
It makes solutions easily move along the gradient with the current fastest critical path depth shortening or area reduction.
Finally, traditional optimizers fall into local optima \cite{zheng2023mitigating, zhao2021multi}.
\begin{figure*}[ht!]
    \centering
    \includegraphics[width=.98\linewidth]{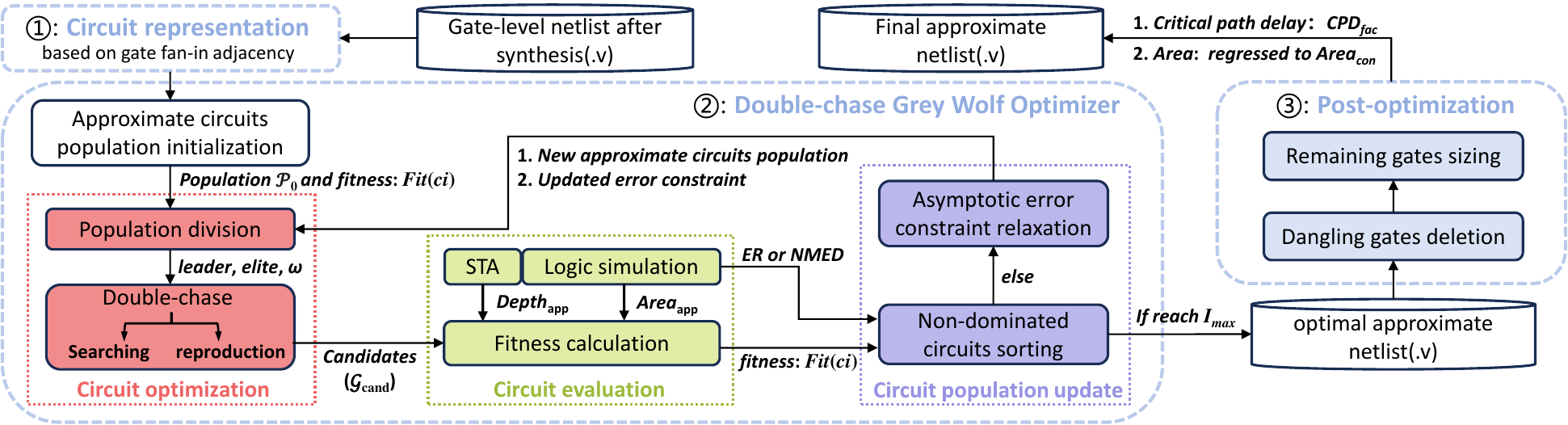}
    \caption{The overall flow of our timing-driven approximate logic synthesis framework based on double-chase grey wolf optimizer.}
    \label{Overflow}
\vspace{-0.1in}
\end{figure*}

In this work, we propose a timing-driven approximate logic synthesis framework. 
As shown in \Cref{Overflow}, the framework is composed of three steps, including circuit representation, the double-chase grey wolf optimizer (DCGWO), and post-optimization.
Firstly, adjacency lists are constructed based solely on gate fan-in relationships to enable fast circuit structure storage and LACs application.
Then, DCGWO efficiently optimizes both critical path depth and area under error constraints.
Subsequently, post-optimization under area constraints converts the area reduction into further timing optimization.
Our contributions are summarized as follows:

\begin{itemize}
\item We propose a framework for deeply exploiting timing improvement inherent in the reduction of critical path depth and the enhancement of gate drive strength.
\item We represent accurate and approximate circuits based on gate fan-in adjacency lists to improve storage efficiency and accelerate timing optimization.
\item We present a DCGWO to effectively select approximate actions for reducing critical path depth and area.
Building upon traditional GWO, it divides the generated approximate circuit population into finer hierarchies and precisely formulates appropriate optimization gradients for each hierarchy, improving the efficiency in finding the global optimal approximate circuit.
\item The experimental results demonstrate that our framework achieves an average $27.13\%$ and $38.54\%$ critical path delay reduction respectively, under a $5\%$ error rate constraint and under a $2.44\%$ normalized mean error distance constraint, outperforming the state-of-the-art method.
\end{itemize}

\section{Preliminaries}

\subsection{Error Metrics}
\label{sec:metrics}
The error metrics used in our framework are error rate ($ER$) and normalized mean error distance ($NMED$). $ER$ can be used to measure the error of random/control circuits, while $NMED$ can evaluate the error of arithmetic circuits. 

For a circuit with $m$ primary inputs and $n$ primary outputs, we assume the probability of input vector $I_{i}$ occurring is $p_{i}$, where $1 \leq i \leq  2^m$. 
In this case, $ER$ is the probability that the approximate circuit output differs from the accurate circuit output, calculated by \Cref{eq:ER}, where $O_{i}^{\text {app}}$ and $O_{i}^{\text {ori}}$ are output vectors of the approximate circuit and accurate circuit for input vector $I_{i}$.
\begin{equation}
ER = \sum \limits_{i=1}^{2^m} (O_{i}^{\text {app}}\ne O_{i}^{\text {ori}})\times p_{i}.
\label{eq:ER}
\end{equation}

Error distance is the difference between approximate circuit output value $V_{i}^{\text {app}}$ and accurate circuit output value $V_{i}^{\text {ori}}$ under input vector $I_{i}$.
$NMED$ is the mean error distance normalized by the maximum output value, defined in \Cref{eq:NMED}. 
\begin{equation}
NMED = \sum \limits_{i=1}^{2^m}
\frac{\left | V_{i}^{\text {ori}} - V_{i}^{\text {app}} \right |}{2^n - 1}\times p_{i}.
\label{eq:NMED}
\end{equation}


\subsection{Problem Formulation}
\label{ProF}
As introduced in \Cref{sec:Intro}, optimizing both critical path depth and area can effectively exploit the potential timing improvement inherent in critical path depth reduction and the enhancement of gate drive strength.
Thus, the timing-driven ALS problem can be formulated as follows:
\begin{myproblem}[Timing-driven ALS]
\label{pro:1}
\textit{Given a post-synthesis netlist of the accurate circuit with timing, area, and logic information, use an approximate optimizer simultaneously optimizing both critical path depth and area under error constraints to generate the final approximate circuit with maximum critical path delay reduction.}
\end{myproblem}

\section{Proposed Framework}
\label{Optimizer Overflow}
The overall flow of our proposed framework is given in \Cref{Overflow}.
In step \textcircled{\footnotesize{$1$}}, the accurate gate-level netlist is represented by gate fan-in adjacency lists.
In step \textcircled{\footnotesize{$2$}}, double-chase grey wolf optimizer (DCGWO) simultaneously optimizes critical path depth and area under error constraints.
It can efficiently search for optimal approximate circuit through iterative circuit optimization, evaluation and circuit population update. 
In step \textcircled{\footnotesize{$3$}}, by performing dangling gates deletion and remaining gates sizing under area constraint $\text{Area}_{con}$ on the generated optimal approximate circuit, the final approximate netlist with maximum critical path delay reduction can be obtained.


\subsection{Circuit Representation}
\label{CR}
\begin{figure}[t]
    \centering
    \includegraphics[width=1\linewidth]{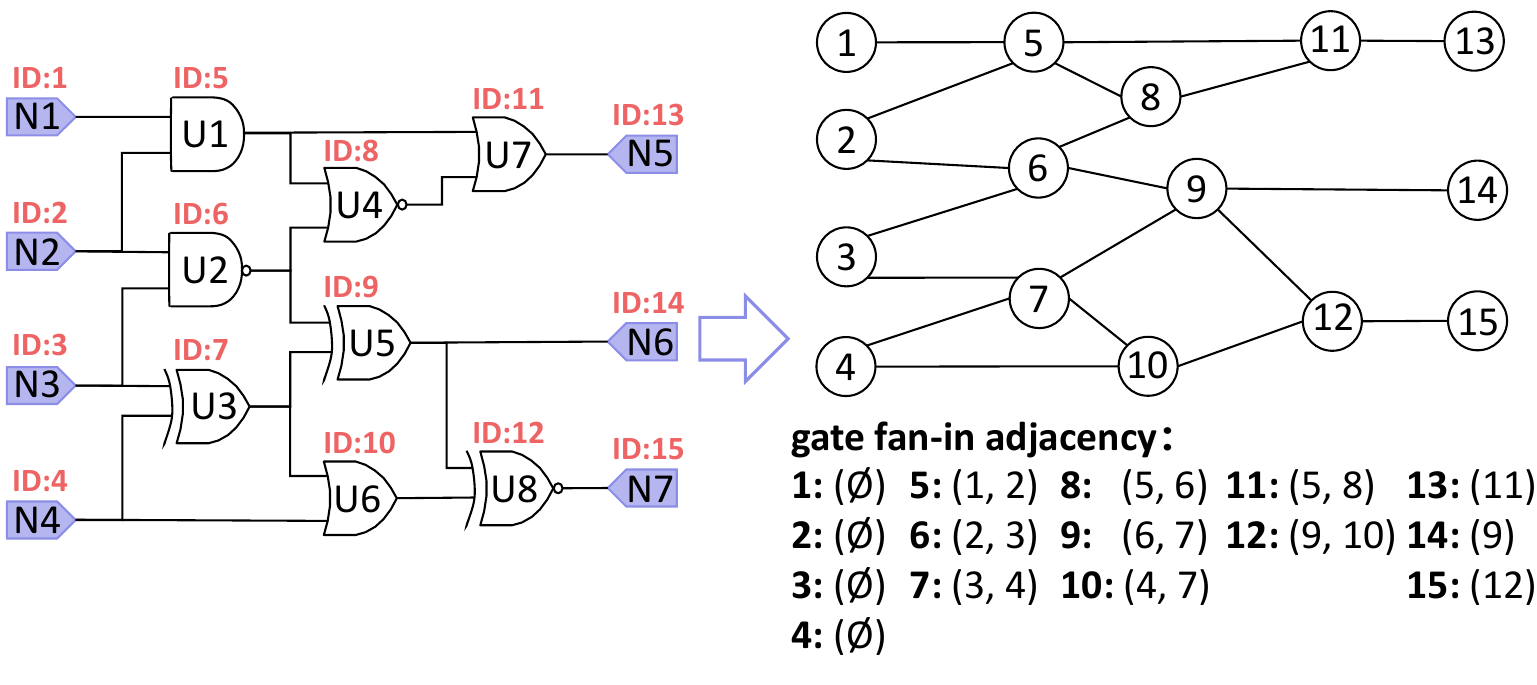}
    \caption{Circuit representation based on gate fan-in adjacency.}
    \label{adj}
\vspace{0.4in}
\end{figure}

We construct adjacency lists storing the circuit structure based solely on fan-in relationships between gates.
By discarding all wire information, the LACs used in our framework, including wire-by-wire \cite{venkataramani2013substitute} and wire-by-constant \cite{schlachter2017design} replacements (shown in \Cref{gca}), can be easily implemented by changing the gate fan-in adjacency.
This operation mode enables us to efficiently assess the impact of LACs and generate corresponding approximate netlist.
To check for circuit loop violations, we further label each gate with a unique integer ID.
\Cref{adj} shows an example of circuit representation, the circuit on the left is stored as fan-in adjacency lists on the right.

To accommodate this circuit representation method, we update the related definitions of above two LACs: the gate to be changed is called target gate, while the gate used for change (constant `0$/$1' are also treated as gates) is called switch gate.

\subsection{Double-chase Grey Wolf Optimizer}
\label{ACG}
In DCGWO, we first generate the initial approximate circuits population $\mathcal{P}_{0}$: $\{ \forall c_i \in \mathcal{P}_{0}\}$ by performing LACs on randomly selected target gates of the accurate circuit.
Each approximate circuit in $\mathcal{P}_{0}$ is evaluated for fitness (defined in \cref{fitness} as function $Fit$), which is composed of critical path depth and area.
Circuits with higher fitness values indicate better quality. 

\begin{figure}[t]
    \centering
    \includegraphics[width=1\linewidth]{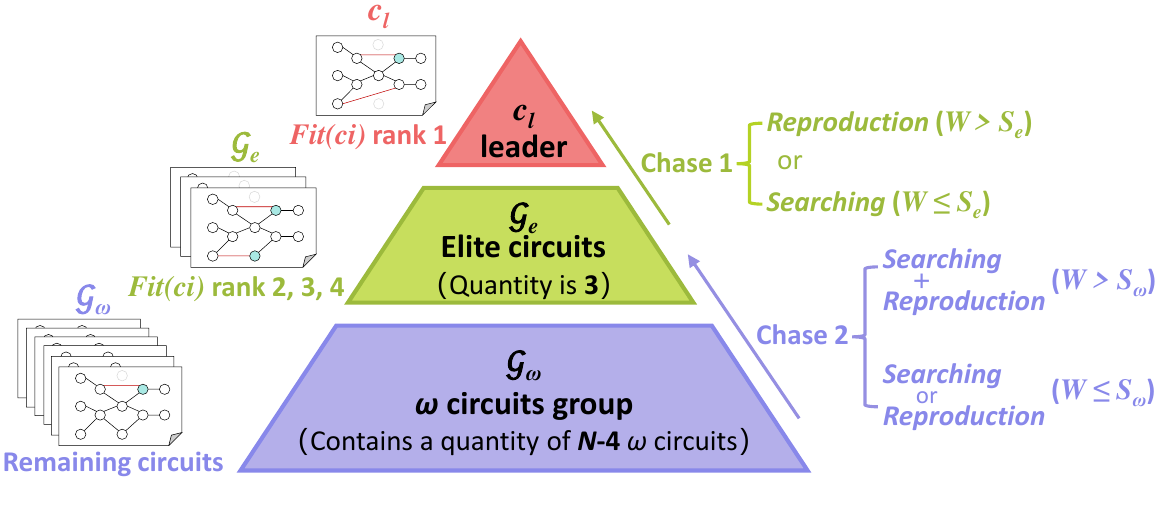}
    \caption{Population division. Population is divided into leader $c_l$, elite circuits $\mathcal{G}_{e}$, and $\omega$ circuits group $\mathcal{G}_\omega$ based on fitness, with each hierarchy engaging in distinct chase operations.}
    \label{population division}
\vspace{-0.04in}
\end{figure}

\minisection{Circuit Optimization}
As shown in \Cref{Overflow}, in each iteration, we perform the double-chase strategy to optimize approximate circuits in the population.
The preliminary work for double-chase involves the population division shown in \Cref{population division}.
It divides the population into leader circuit $c_l$, elite circuits $\mathcal{G}_{e}$, and $\omega$ circuits group $\mathcal{G}_\omega$ based on their fitness values.
Specifically, the leader circuit $c_l$ is the approximate circuit with the highest fitness.
It guides the elite circuits with fitness ranks 2, 3, and 4 in Chase 1.
The elite circuits guide $\omega$ circuits group in Chase 2.
For Chase 1 and 2, we design two approximate actions: \textbf{circuit searching} and \textbf{circuit reproduction}. They are used alternately to generate new approximate circuits along suitable optimization gradients.


The circuit searching essentially uses wire-by-wire and wire-by-constant to shorten critical paths. 
Specifically, we first use PrimeTime \cite{primetime} to obtain the critical paths with maximum propagation time from primary input (PI) to primary output (PO).
Then, for each critical path, all gates on it are stored in the targets set $\mathcal{T}_c$ and undergo uniform (0,1) distribution sampling.
All fan-ins of sampled gates with a probability greater than 0.5 are also stored in $\mathcal{T}_c$.
The target gate is randomly selected from $\mathcal{T}_c$.
To limit introduced error, switch gate is selected based on similarities, i.e., the percentage of cycles when output of target gate holds the same value with output of each gate in its transitive fan-in (TFI) or the constant logic value `0', `1'.
The gate or constant logic value with the highest similarity is selected to substitute the target gate.

\begin{figure*}[ht!]
    \centering
    \includegraphics[width=1\linewidth]{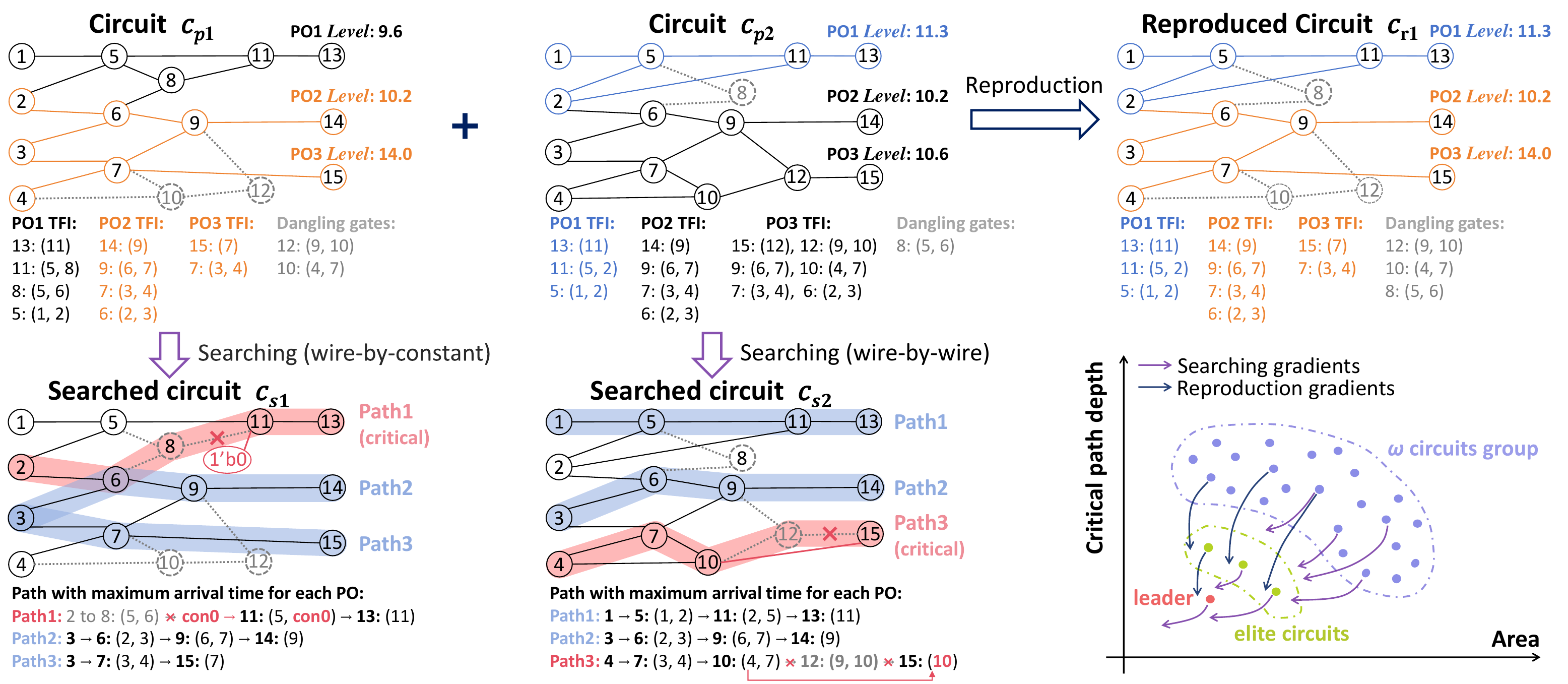}
    \caption{Illustrations of the circuit searching, circuit reproduction, and optimization gradients guided by them in double-chase.}
    \label{ReP}
\vspace{-0.1in}
\end{figure*}

\Cref{ReP} shows circuit searching examples.
For obtaining $c_{s1}$ from $c_{p1}$ through wire-by-constant searching, Path1 is the critical path. Thus we select ID8 gate (outputs: 14 cycles of `0' and 2 cycles of `1') as the target gate, and constant logic value `0' with the highest similarity 0.875 as the switch gate.
In this case, the fan-in adjacency of the ID11 gate is changed from (5, 8) to (5, con0), greatly decreasing the Path1 depth.
Similarly, for obtaining $c_{s2}$ from $c_{p2}$, the fan-in adjacency of ID15 PO is changed from 12 to 10, decreasing the Path3 depth.

Inspired by a crossover in genetic algorithm \cite{deb2002fast}, circuit reproduction is designed to aggregate well-optimized path sets with low errors from two selected approximate circuits, generating a reproduced circuit with better quality.
Specifically, we first divide each selected circuit according to the POs and corresponding TFI.
Then, for each PO, we use its maximum arrival time $T_a$ and the error generated on it $Error$ to form the PO-TFI pair evaluation function $Level$ in \Cref{level_calc}, where $w_{\text {t}}$ and $w_{\text {e}}$ are the weights of $T_a$ and $Error$ respectively. 
\begin{equation}
Level(PO_i)=w_{\text {t}}\times \frac{1}{T_a(PO_i)} + 
w_{\text {e}}\times \frac{1}{Error(PO_i)}.
\label{level_calc}
\end{equation}

Subsequently, we choose PO-TFI pairs with higher $Level$ from two selected circuits to form the reproduced circuit.
Some gates are shared by different PO-TFI pairs. 
Thus, gates in the reproduced circuit only accept adjacency information from the first write-in.
Taking circuits $c_{p1}$ and $c_{p2}$ in \Cref{ReP} as an example, by comparing their $Level$, we select PO2-TFI, PO3-TFI pairs from $c_{p1}$, and PO1-TFI pair from $c_{p2}$, to form circuit $c_{r1}$.
Since gates with IDs 8, 10 and 12 are not in any PO-TFI pair, to ensure the completeness of $c_{r1}$, their information is selected from $c_{p1}$ and $c_{p2}$ to write in $c_{r1}$.

\Cref{population division} illustrates that approximate circuits at different hierarchies consult their adjacent higher-hierarchy circuits for circuit searching and reproduction.
Therefore, we design the fitness distance $D$, decision parameter $W$ and decision threshold $S$ for both elite circuits $\mathcal{G}_{e}$ and $\omega$ circuits group $\mathcal{G}_\omega$.
$D$ is calculated by \Cref{D}, where $r_c$ is defined as a random value between $[0,2]$.
Since $\mathcal{G}_{e}$ reference the leader circuit $c_l$ for Chase 1, $D$ for elite circuits in $\mathcal{G}_{e}$ include the fitness of leader circuit $Fit(c_l)$.
Similarly, $\mathcal{G}_\omega$ reference $\mathcal{G}_e$ for Chase 2.
Thus, $D$ for $\omega$ circuits in $\mathcal{G}_\omega$ include the average fitness of elite circuits in $\mathcal{G}_e$.
$W$ provides a dynamic correction to $D$ by adding the encircling coefficient $A$.
\begin{align}
\label{D}
	D(c_i) =
	\begin{cases}
		r_c\times Fit(c_l)-Fit(c_i) \hfill \forall c_i\in \mathcal{G}_{e}\\
		\frac{r_c}{3} \sum_{c_j\in \mathcal{G}_e} Fit\left ( c_j \right )-Fit(c_i) \quad \hfill \forall c_i\in \mathcal{G}_{\omega}\\
	\end{cases},
\end{align}

\begin{equation}
W(c_i)=A\times D(c_i),
\label{W}
\end{equation}
where $A$ is calculated based on the scaling factor $a$ as:
\begin{equation}
A=(2\times r_1-1)\times a,
\label{A}
\end{equation}
where $r_1$ is a random value between $[0,1]$. 
The scaling factor $a$ balances the global search and local convergence of the population during the iterative process.
As shown in \Cref{a}, $a$ decreases with the increase of iteration $iter$ until $iter$ reaches the upper limit of iteration $I_{\max}$.
\begin{equation}
a=2-\frac{2\times iter}{I_{\max}}.
\label{a}
\end{equation}



As shown in \Cref{population division}, the approximate actions are decided by the relationship between decision parameter $W$ and decision thresholds $S$. 
Decision thresholds used for $\mathcal{G}_{e}$ and $\mathcal{G}_\omega$ are $S_e$ and $S_{\omega}$, respectively.
For circuit $c_i$ in $\mathcal{G}_{e}$, if $W(c_i)>S_e$, it executes circuit reproduction with another circuit of superior fitness to generate a reproduced circuit.
Otherwise, it uses circuit searching to reduce its critical path depth and area.
Meanwhile, for circuit $c_i$ in $\mathcal{G}_{\omega}$, if $W(c_i)>S_\omega$, it performs both circuit searching and reproduction.
Otherwise, it randomly selects either circuit searching or reproduction.
When the double-chase is completed, the leader $c_l$ conducts circuit searching to ensure its variability.
Then, approximate circuits before and after double-chase are stored in the candidates' group $\mathcal{G}_{cand}$ for further evaluation and update. 

The lower right corner of \Cref{ReP} demonstrates that the double-chase strategy can effectively guide the entire population to move along the appropriate gradient with simultaneous critical path depth and area reductions.



\minisection{Circuit Fitness Evaluation}
\label{fitness}
The fitness function is composed of two optimization objectives: critical path depth and area.
The depth-related information, including the maximum critical path depth of each approximate circuit $Depth_{\text{app}}$ and the depth of the longest path in corresponding accurate circuit $Depth_{\text{ori}}$, are obtained through static timing analysis using PrimeTime \cite{primetime}.
Since circuit searching and reproduction change the connection relationship between gates, some gates become dangling due to their inability to connect to any PO.
Therefore, the area of each approximate circuit $Area_{\text{app}}$ is the area of accurate circuit $Area_{\text {ori}}$ minus the area of these dangling gates.

The fitness function $Fit$ of approximate circuit $c_i$ is defined in \Cref{fit_calc}, where $w_{\text {d}}$ and $w_{\text {a}}=1-w_{\text {d}}$ respectively denote the weights assigned to the critical path depth and area. 
Circuits with higher fitness values indicate better quality.
\begin{equation}
Fit(c_i)=w_{\text {d}}\times \frac{Depth_{\text {ori}}(c_i)}{Depth_{\text {app}}(c_i)} + 
w_{\text {a}}\times \frac{Area_{\text {ori}}(c_i)}{Area_{\text {app}}(c_i)}.
\label{fit_calc}
\end{equation}

\minisection{Circuit Population Update}
\label{pu}
To select high-quality approximate circuits under error constraints, we perform non-dominated sorting \cite{zhang2014efficient} on the evaluated candidates' group $\mathcal{G}_{cand}$.
It is achieved based on Pareto dominance between circuits determined by two functions: depth function $f_d=\frac{Depth_{\text {ori}}}{Depth_{\text {app}}}$ and area function $f_a=\frac{Area_{\text {ori}}}{Area_{\text {app}}}$.
Firstly, we remove circuits exceeding the specified error constraint from $\mathcal{G}_{cand}$. 
Then, we maintain the dominated list $\mathcal{L}_{\text{d}}$ for each remaining circuit. 
For approximate circuits $c_i$ and $c_j$,  if $c_i$ is not inferior to $c_j$ in two functions, and is superior in at least one of them, then $c_i$ dominates $c_j$ and is added to $\mathcal{L}_{\text{d}}$
of $c_j$.
Circuits with empty $\mathcal{L}_{\text{d}}$ are considered Pareto-optimal circuits. We place them into the 0-ranked Pareto set while removing them from $\mathcal{G}_{cand}$ and the $\mathcal{L}_{\text{d}}$ of other circuits.
Subsequently, new Pareto-optimal circuits with empty $\mathcal{L}_{\text{d}}$ emerge, forming the 1-ranked Pareto set, and undergo the same removal process. 
This will repeat until $\mathcal{G}_{cand}$ is empty.

We further calculate the crowding distance $Dist$ of approximate circuits in each Pareto set.
With higher $Dist$, circuits are less likely to overlap in the objective function space, resulting in better optimization efficiency.
In the $k$-ranked Pareto set, approximate circuits are sorted separately based on $f_d$ and $f_a$.
The $Dist$ of the circuits at the beginning and end of these two sorted lists are set to $+\infty$.
For approximate circuit $c_i$, circuits adjacent to $c_i$ in each sorted list are $c_{i-1}$ and $c_{i+1}$.
In this case, $Dist$ is calculated by \Cref{D_crowd}.
\begin{equation}
Dist(c_i)=\sum_{x=d,a}\frac{f_x(c_{i-1})-f_x{(c_{i+1})}}{\max_k(f_x)-\min_k(f_x)}.
\label{D_crowd}
\end{equation}

Based on Pareto set partition and crowding distance calculation, we sort the approximate circuits within each Pareto set in descending order of their $Dist$.
Subsequently, starting from the 0-ranked Pareto set, we sequentially select ${N}$ approximate circuits to form a new population for the next iteration.

After the non-dominated circuits sorting, the asymptotic error constraint relaxation is employed.
We design a quadratic function scheme (i.e., $Error^{iter}_{cons.} = b\times iter^2 + Error^{0}_{cons.}$) to gradually increase the error constraint $Error^{iter}_{cons.}$ as the iteration $iter$ rises, ultimately relaxing it to the user-specified maximum error constraint by setting appropriate empirical parameter $b$.
This operation prevents the population from quickly moving to the maximum error constraint boundary and getting trapped in local optima.

\subsection{Post-Optimization}
\label{GS}

Post-optimization is performed on the optimal approximate circuit generated by DCGWO.
It can further convert the area reductions into timing performance improvements by enhancing the drive strength of gates.

We first delete dangling gates produced by circuit searching and reproduction from the optimal approximate circuit.
In this process, we traverse the entire circuit, identifying and removing gates with empty transitive fan-out (TFO).
For each fan-in of the removed gates, we similarly perform identification and removal operations until no gates with empty TFO remain.
Subsequently, for the processed optimal approximate circuit, we use Design Compiler \cite{designcompiler} to resize its remaining gates without adjusting any circuit structure under area constraints $\text{Area}_{con}$.
Consequently, the final approximate circuits with minimum critical path delay $CPD_{fac}$ are obtained.

\section{Experimental Results}

Our proposed framework is implemented in Python.
We set up the experimental environment on the Linux machine with 32 cores and 4 NVIDIA Tesla V100 GPUs in parallel with 128GB memory. 
The benchmarks listed in \Cref{tab:benchmarks} are from ISCAS'85 \cite{hansen1999unveiling} and EPFL \cite{EPFL}. 
Each circuit is synthesized into gate-level netlist by Design Compiler \cite{designcompiler} under TSMC 28nm technology.
Among these benchmarks, random/control circuits are optimized under $ER$ constraints, while arithmetic circuits are optimized under $NMED$ constraints. 
For the generated approximate circuits, their timing-related information is obtained through static timing analysis performed by PrimeTime \cite{primetime}.
The circuit error and the similarities between outputs of gates are obtained using VECBEE based on Monte Carlo simulation \cite{su2022vecbee}. 
By setting the number of sampled input vectors to $1\times 10^{5}$, this method can achieve fast error and similarities evaluation with nearly no deviation.

\subsection{Parameter Setting}
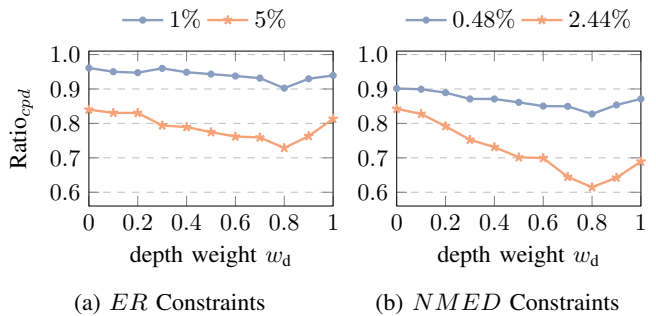
\begin{figure}[t!]
\flushleft
\hspace{-.21in}
\subfloat[$ER$ Constraints]{\pgfplotsset{compat=newest}

\pgfplotsset{
	width =0.28\textwidth,
	height=0.21\textwidth
}

\begin{tikzpicture}[scale=0.9]
	\begin{axis}[
		samples=1000,
		xmax=1, xmin=0,
		ymax=1.01, ymin=0.56,
		xtick={0,0.2,0.4,0.6,0.8,1},
		ytick={0.6,0.7,0.8,0.9,1.0},
		ymajorgrids=true,
		grid style=dashed,
		scaled ticks=false,
		legend pos=north east,
		xticklabel style={/pgf/number format/fixed},
		yticklabel style={/pgf/number format/.cd, fixed, fixed zerofill, precision=1, /tikz/.cd},
		xlabel={depth weight $w_{\text{d}}$},
		ylabel={$\text{Ratio}_{cpd}$},
		xlabel near ticks,
		ylabel near ticks,
		legend style={
			draw=none,
			at={(0.1,1.2)},
			anchor=west,
			legend columns=-1,
		}
		]
		
		\pgfplotstableset{
			create on use/x rel/.style={
				create col/expr={
					\thisrow{0}
				}
			},
			create on use/y rel/.style={
				create col/expr={
					\thisrow{1}
				}
			}
		}

		\addplot +[myblue1, line width=1pt, mark=*, mark options={scale=0.5, fill=myblue1},text mark as node=true ] table [x ={ratio}, y={cpd}] {er1.dat};
		\addplot +[myorange1, line width=1pt, mark=star, mark options={scale=1.0, fill=myorange1},text mark as node=true ] table [x ={ratio}, y={cpd}] {er5.dat};	
  \legend{$1\%$, $5\%$}

	\end{axis}

\end{tikzpicture} \label{fig:weightER}}
\hspace{-.23in}
\subfloat[$NMED$ Constraints]{\pgfplotsset{compat=newest}

\pgfplotsset{
	width =0.28\textwidth,
	height=0.21\textwidth
}

\begin{tikzpicture}[scale=0.9]
	\begin{axis}[
		samples=1000,
		xmax=1, xmin=0,
		ymax=1.01, ymin=0.56,
		xtick={0,0.2,0.4,0.6,0.8,1},
		ytick={0.6,0.7,0.8,0.9,1.0},
		ymajorgrids=true,
		grid style=dashed,
		scaled ticks=false,
		legend pos=north east,
		xticklabel style={/pgf/number format/fixed},
		yticklabel style={/pgf/number format/.cd, fixed, fixed zerofill, precision=1, /tikz/.cd},
		xlabel={depth weight $w_{\text{d}}$},
		xlabel near ticks,
		legend style={
			draw=none,
			at={(0,1.2)},
			anchor=west,
			legend columns=-1,
		}
		]
		
		\pgfplotstableset{
			create on use/x rel/.style={
				create col/expr={
					\thisrow{0}
				}
			},
			create on use/y rel/.style={
				create col/expr={
					\thisrow{1}
				}
			}
		}

		\addplot +[myblue1, line width=1pt, mark=*, mark options={scale=0.5, fill=myblue1},text mark as node=true ] table [x ={ratio}, y={cpd}] {nmed48.dat};
		\addplot +[myorange1, line width=1pt, mark=star, mark options={scale=1.0, fill=myorange1},text mark as node=true ] table [x ={ratio}, y={cpd}] {nmed244.dat};	
  \legend{$0.48\%$, $2.44\%$ }

	\end{axis}

\end{tikzpicture} \label{fig:weightNMED}}	
\caption{Average critical path delay ratios $\text{Ratio}_{cpd}$ generated by our framework using different depth weight $w_{\text{d}}$ under the tightest and loosest $ER$ and $NMED$ constraints.}
\label{weightchange}
\end{figure}

The parameters of our framework are set as follows.
The population size ${N}$ is 30 and the upper limit of iterations $I_{max}$ is 20.
For PO-TFI pair evaluation function $Level$, $w_{\text{t}}$ is $0.9\times CPD_\text{ori}$ under both error constraints, while $w_{\text{e}}$ is respectively 0.1 and 0.2 under $ER$ and $NMED$ constraints.
For circuit fitness $Fit$, we determine the optimal weights based on \textbf{critical path delay ratios} of the final approximate circuits over the accurate circuits (i.e., $\text{Ratio}_{cpd}=\frac{CPD_{fac}}{CPD_\text{ori}}$). 
\Cref{weightchange} illustrates that the minimum $\text{Ratio}_{cpd}$ are achieved under both the tightest and loosest error constraints when $w_{\text{d}}$ is 0.8 and $w_{\text{a}} = 1-w_{\text{d}}$ is 0.2. 
Therefore, we follow this setting.
\begin{table}[t]
	\centering
	\caption{The benchmark statistics. $CPD_\text{ori}$ (ps) and $Area_\text{ori}$ ($\mu m^2$) respectively represents the \textbf{critical path delay} and area of accurate circuit.} 
\label{tab:benchmarks}
\resizebox{1.0\linewidth}{!}
{
    \Huge 
	\begin{tabular}{|c|lllcc|l|}
		\hline
		Type                       & Circuit    & \#gate & \#PI$/$PO & $CPD_\text{ori}$ & $Area_\text{ori}$  & Description              \\ \hline \hline
		  \multirow{2}{*}{}                   & Cavlc      & 573    & 10$/$11      & 186.35       & 450.31       & Coding Cavlc         \\
                                                & c880       & 322    & 60$/$26      & 185.34       & 177.67        & 8-bit ALU          \\
            Random$/$                           & c1908      & 366    & 33$/$25      & 235.14       & 223.34       & 16-bit SEC/DED circuit         \\
            Control                             & c2670      & 922    & 233$/$140     & 218.40      & 288.71       & 12-bit ALU and controller         \\
            \multirow{3}{*}{}                   & c3540      & 667    & 50$/$22      & 293.09       & 459.42       & 8-bit ALU         \\
                                                & c5315      & 2595   & 178$/$123     & 122.25      & 1129.55      & 9-bit ALU         \\
                                                & c7552      & 1576   & 207$/$108     & 282.13      & 939.33       & 32-bit adder/comparator         \\ \hline
                                                  
            \multirow{8}{*}{Arithmetic}         & Int2float  & 198    & 11$/$7      & 127.02        & 194.63       & int to float converter       \\ 
                                                & Adder16    & 269    & 32$/$17      & 58.92       & 288.41       & 16-bit adder          \\
                                                & Max16      & 154    & 32$/$16      & 131.78       & 91.43        & 16-bit 2-1 max unit          \\
                                                & c6288      & 1641   & 32$/$32      & 847.79       & 687.08       & 16$\times$16 multiplier         \\
                                                & Adder      & 1639   & 256$/$129     & 1394.7      & 495.78       & 128-bit adder            \\
		                                    & Max        & 2940   & 512$/$120    & 2799.8      & 954.03       & 128-bit 4-1 max unit     \\
		                                    & Sin        & 10962  & 24$/$25      & 701.03       & 4367.27      & 24-bit sine unit         \\ 
                                                & Sqrt       & 13542  & 128$/$64     & 67929.3       & 6262.10       & 128-bit square root unit    \\ \hline

	\end{tabular}
}
\vspace{-0.1in}
\end{table}

\begin{table*}[t!]
\centering
\caption{Comparison of performance between our framework and others under $5\%$ $ER$ constraints. All final generated circuits experience post-optimization under area constraints $\text{Area}_{con}$ to convert area reduction into further critical path delay reduction.}
\label{tab:er}
\resizebox{.94\linewidth}{!}
{
\begin{tabular}{|l|c|cc|cc|cc|cc|cc|}
\hline
\multirow{2}{*}{Circuit}   &{$\text{Area}_{con}$}  &\multicolumn{2}{c|}{VECBEE-S \cite{su2022vecbee}} &\multicolumn{2}{c|}{VaACS \cite{balaskas2022variability}}   &\multicolumn{2}{c|}{HEDALS\cite{meng2023hedals}}   &\multicolumn{2}{c|}{GWO (single-chase)}   &\multicolumn{2}{c|}{Ours} \\ 
                           &($\mu m^2$)          & $\text{Ratio}_{cpd}$         & runtime(s)        & $\text{Ratio}_{cpd}$          & runtime(s)     & $\text{Ratio}_{cpd}$          & runtime(s)      & $\text{Ratio}_{cpd}$     & runtime(s)   & $\text{Ratio}_{cpd}$          & runtime(s) \\ \hline \hline
Cavlc         &450.00  & 0.9219  & \textbf{60.03}  & 0.8745  & 356.89  &0.9071   &194.43   &0.8963   &407.25   & \textbf{0.8602}  & 310.42 \\
c880          &177.00  & 0.9026  & \textbf{43.11}  & 0.9221  & 227.13  &0.8913   &104.00   &0.9183   &201.51   & \textbf{0.8399}  & 193.86  \\
c1908         &223.00  & 0.8679  & \textbf{65.32}  & 0.5166  & 235.68  &\textbf{0.3372}   &310.42   &0.5021   &307.56   & 0.3865  & 202.79  \\
c2670         &288.00  & 0.6708 & 308.16 & 0.8101 & 477.92 &0.7589   &\textbf{250.28}   &0.7703   &313.99   & \textbf{0.6314} & 339.63 \\
c3540         &459.00  & 0.9670 & 391.42 & 0.9729 & 435.26 & 0.9203   &373.26   &0.9224   &479.88   & \textbf{0.8732} & \textbf{324.59} \\
c5315         &1129.00 & 0.9113 & 1857.32 & 0.8599 & 1963.55 &0.8270   &1662.08   &0.8165   &1655.07   & \textbf{0.8034} & \textbf{1449.37} \\
c7552         &939.00  & 0.9262 & 1726.27 & 0.9133  & 1336.64 &0.7391   &1315.85   &0.8877   &1420.32   & \textbf{0.7063}  & \textbf{1279.18} \\ \hline \hline 
Average       &523.57  & 0.8811 & 635.94 & 0.8385 & 719.01 &0.7687   &601.47   &0.8162   &683.65   & \textbf{0.7287} & \textbf{585.69} \\ \hline
\end{tabular}
}
\vspace{-0.01in}
\end{table*}

\begin{table*}[t!]
\centering
\caption{Comparison of performance between our framework and others under $2.44\%$ $NMED$ constraints. All final generated circuits experience post-optimization under $\text{Area}_{con}$ to convert area reduction into further critical path delay reduction.}
\label{tab:nmed}
\resizebox{.94\linewidth}{!}
{ 
\begin{tabular}{|l|c|cc|cc|cc|cc|cc|}
\hline
\multirow{2}{*}{Circuit}    &{$\text{Area}_{con}$}   & \multicolumn{2}{c|}{VECBEE-S \cite{su2022vecbee}} & \multicolumn{2}{c|}{VaACS \cite{balaskas2022variability}}    &\multicolumn{2}{c|}{HEDALS\cite{meng2023hedals}}   &\multicolumn{2}{c|}{GWO (single-chase)}   & \multicolumn{2}{c|}{Ours} \\ 
                                &($\mu m^2$)          & $\text{Ratio}_{cpd}$         & runtime(s)        & $\text{Ratio}_{cpd}$          & runtime(s)     & $\text{Ratio}_{cpd}$          & runtime(s)      & $\text{Ratio}_{cpd}$     & runtime(s)   & $\text{Ratio}_{cpd}$          & runtime(s) \\ \hline \hline
Int2float       &194.00  & 0.9331  & 71.23 & 0.5047  & 151.73 &0.7649 &\textbf{32.68} &0.6010 &178.30 & \textbf{0.4496}  & 132.12 \\
Adder16         &288.00  & 0.9973  & 67.20  & 0.5295  & 173.85  &0.4513 &\textbf{47.30} &0.5216 &189.01 & \textbf{0.4275}  & 167.03  \\
Max16           &91.00   & 0.7087  & \textbf{93.17} & 0.4209 & 189.73 &0.4470 &105.97 &0.3928 &277.38 & \textbf{0.3708} & 208.55 \\
c6288           &687.00  & 0.9663  & 4410.29  & 0.8696  & 3279.62  &\textbf{0.6368} &2563.41 &0.9079 &2991.00 & 0.8313  & \textbf{2103.88}  \\
Adder           &495.00  & 0.7814  & 1697.37  & 0.8133 & 2083.15  &0.7110 &1362.70 &0.8008 &1550.03 & \textbf{0.6917}  & \textbf{1193.71}  \\
Max             &954.00  & 0.8809  & 2600.78 & 0.8933 & 3397.50 &0.8355 &2992.08 &0.7517 &3121.44 & \textbf{0.6799} & \textbf{2035.62} \\
Sin             &4367.00 & 0.9187  & 5391.68  & 0.8326  & 3872.31  &0.7945 &3380.52 &0.8722 &4392.77 & \textbf{0.7603}  & \textbf{3176.46}  \\
Sqrt            &6262.00 & 0.7993  & 33117.12  & 0.8011  & 20160.76  &0.7437 &11242.29 &0.7803 &17894.50 & \textbf{0.7058}  & \textbf{9950.11}  \\ \hline \hline
Average         &1667.25 & 0.8732   & 5931.11  & 0.7081  & 4163.58 &0.6731 &2715.87 &0.7035 &3824.30 & \textbf{0.6146} & \textbf{2370.94} \\ \hline
\end{tabular}
}
\vspace{-0.16in}
\end{table*}

\subsection{Optimization Performance}
Since our framework focuses on timing optimization, we compare the performance of our framework, including final \textbf{critical path delay ratios} $\text{Ratio}_{cpd}=\frac{CPD_{fac}}{CPD_\text{ori}}$ and \textbf{runtime}, with: (1) area-driven methods: VECBEE-SASIMI \cite{su2022vecbee}; (2) depth-driven methods: VaACS \cite{balaskas2022variability}, HEDALS \cite{meng2023hedals}; (3) traditional GWO (single-chase). 
Approximate circuits generated by these works experience post-optimization (in \Cref{GS}) under \textbf{area constraints} $\text{Area}_{con}$ to convert area reduction into further critical path delay reduction by Design Compiler {\cite{designcompiler}}.

For random/control circuits, the performance comparison of all works under the loosest $5\%$ $ER$ constraint is detailed in \Cref{tab:er}. 
According to the comparison results, by setting the same area constraints, our framework maximizes the average critical path delay reduction to $27.13\%$ with shorter runtime under the $5\%$ $ER$ constraint.
Similarly, for arithmetic circuits, the performance comparison of all works under the loosest $2.44\%$ $NMED$ constraint is detailed in \Cref{tab:nmed}.
The comparison results indicate that our framework maximizes the average critical path delay reduction to $38.54\%$ with shorter runtime under the $2.44\%$ $NMED$ constraint.

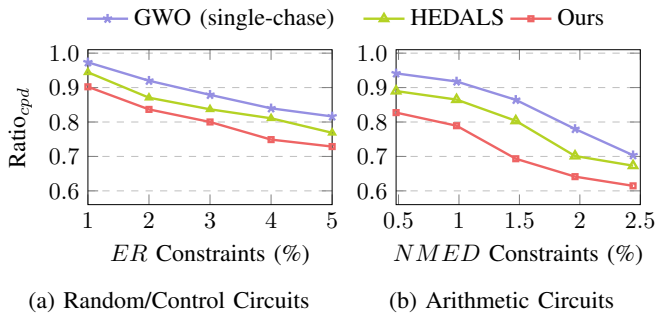
\begin{figure}[t!]
\flushleft
\hspace{-.2in}
\subfloat[Random/Control Circuits]{\pgfplotsset{compat=newest}

\pgfplotsset{
	width =0.28\textwidth,
	height=0.21\textwidth
}

\begin{tikzpicture}[scale=0.9]
	\begin{axis}[
		samples=1000,
		xmax=5.0, xmin=1.0,
		ymax=1.01, ymin=0.56,
		xtick={1.0,2.0,3.0,4.0,5.0},
		ytick={0.6, 0.7,0.8,0.9,1.0},
		ymajorgrids=true,
		grid style=dashed,
		scaled ticks=false,
		legend pos=north east,
		xticklabel style={/pgf/number format/fixed},
		yticklabel style={/pgf/number format/.cd, fixed, fixed zerofill, precision=1, /tikz/.cd},
		xlabel={$ER$ Constraints (\%)},
		ylabel={$\text{Ratio}_{cpd}$},
		xlabel near ticks,
		ylabel near ticks,
		legend style={
			draw=none,
			at={(-0.04,1.2)},
			anchor=west,
			legend columns=-1,
		}
		]
		
		\pgfplotstableset{
			create on use/x rel/.style={
				create col/expr={
					\thisrow{0}
				}
			},
			create on use/y rel/.style={
				create col/expr={
					\thisrow{1}
				}
			}
		}

		\addplot +[mygreen, line width=1pt, mark=triangle, mark options={scale=0.7, fill=mygreen},text mark as node=true ] table [x ={er}, y={cpd}] {cpd-HEDALS.dat};	
		\addplot +[myred, line width=1pt, mark=square, mark options={scale=0.5, fill=myred},text mark as node=true] table [x ={er}, y={cpd}] {cpdOurs.dat};
            \addplot +[mypurple, line width=1pt, mark=star, mark options={scale=1.0, fill=mypurple},text mark as node=true ] table [x ={er}, y={cpd}] {cpdGWO-S.dat};
		\legend{, ,GWO (single-chase)}

	\end{axis}

\end{tikzpicture} \label{fig:ERcon}}
\hspace{-.23in}
\subfloat[Arithmetic Circuits]{\pgfplotsset{compat=newest}

\pgfplotsset{
	width =0.28\textwidth,
	height=0.21\textwidth
}

\begin{tikzpicture}[scale=0.9]
	\begin{axis}[
		samples=1000,
		xmax=2.5, xmin=0.48,
		ymax=1.01, ymin=0.56,
		xtick={0.5,1.0,1.5,2.0,2.5},
		ytick={0.6,0.7,0.8,0.9,1.0},
		ymajorgrids=true,
		grid style=dashed,
		scaled ticks=false,
		legend pos=north east,
		xticklabel style={/pgf/number format/fixed},
		yticklabel style={/pgf/number format/.cd, fixed, fixed zerofill, precision=1, /tikz/.cd},
		xlabel={$NMED$ Constraints (\%)},
		xlabel near ticks,
		legend style={
			draw=none,
			at={(-0.17,1.2)},
			anchor=west,
			legend columns=-1,
		}
		]
		
		\pgfplotstableset{
			create on use/x rel/.style={
				create col/expr={
					\thisrow{0}
				}
			},
			create on use/y rel/.style={
				create col/expr={
					\thisrow{1}
				}
			}
		}

		\addplot +[mygreen, line width=1pt, mark=triangle, mark options={scale=1.0, fill=black},text mark as node=true ] table [x ={er}, y={cpd}] {cpd-HEDALS2.dat};	
		\addplot +[myred, line width=1pt, mark=square, mark options={scale=0.5, fill=mymiddle},text mark as node=true] table [x ={er}, y={cpd}] {cpdOurs2.dat};
            \addplot +[mypurple, line width=1pt, mark=star, mark options={scale=1.0, fill=mypurple},text mark as node=true ] table [x ={er}, y={cpd}] {cpdGWO-S2.dat};
  \legend{HEDALS,Ours}

	\end{axis}

\end{tikzpicture} \label{fig:NMEDcon}}	
\caption{Average critical path delay ratios $\text{Ratio}_{cpd}$ generated by our framework, HEDALS \cite{meng2023hedals} and traditional GWO under different $ER$ and $NMED$ constraints.}
\label{underconstraint}
\vspace{-0.02in}
\end{figure}

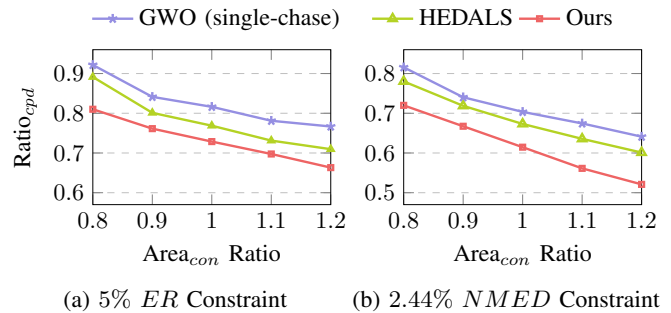
\begin{figure}[t!]
\flushleft
\hspace{-.2in}
\subfloat[$5\%$ $ER$ Constraint]{\pgfplotsset{compat=newest}

\pgfplotsset{
	width =0.275\textwidth,
	height=0.21\textwidth
}

\begin{tikzpicture}[scale=0.9]
	\begin{axis}[
		samples=1000,
		xmax=1.2, xmin=0.8,
		ymax=0.96, ymin=0.57,
		xtick={0.8,0.9,1.0,1.1,1.2},
		ytick={0.6,0.7,0.8,0.9},
		ymajorgrids=true,
		grid style=dashed,
		scaled ticks=false,
		legend pos=north east,
		xticklabel style={/pgf/number format/fixed},
		yticklabel style={/pgf/number format/.cd, fixed, fixed zerofill, precision=1, /tikz/.cd},
		xlabel={$\text{Area}_{con}$ Ratio},
		ylabel={$\text{Ratio}_{cpd}$},
		xlabel near ticks,
		ylabel near ticks,
		legend style={
			draw=none,
			at={(-0.04,1.2)},
			anchor=west,
			legend columns=-1,
		}
		]
		
		\pgfplotstableset{
			create on use/x rel/.style={
				create col/expr={
					\thisrow{0}
				}
			},
			create on use/y rel/.style={
				create col/expr={
					\thisrow{1}
				}
			}
		}

		\addplot +[mygreen, line width=1pt, mark=triangle, mark options={scale=0.7, fill=mygreen},text mark as node=true ] table [x ={a}, y={cpd}] {cpd-HEDALSA.dat};	
		\addplot +[myred, line width=1pt, mark=square, mark options={scale=0.5, fill=myred},text mark as node=true] table [x ={a}, y={cpd}] {cpdOursA.dat};
            \addplot +[mypurple, line width=1pt, mark=star, mark options={scale=1.0, fill=mypurple},text mark as node=true ] table [x ={a}, y={cpd}] {cpdGWO-A.dat};
		\legend{, ,GWO (single-chase)}

	\end{axis}

\end{tikzpicture} \label{fig:AreaconER}}
\hspace{-.23in}
\subfloat[$2.44\%$ $NMED$ Constraint]{\pgfplotsset{compat=newest}

\pgfplotsset{
	width =0.275\textwidth,
	height=0.21\textwidth
}

\begin{tikzpicture}[scale=0.9]
	\begin{axis}[
		samples=1000,
		xmax=1.2, xmin=0.8,
		ymax=0.86, ymin=0.47,
		xtick={0.8,0.9,1.0,1.1,1.2},
		ytick={0.5,0.6,0.7,0.8},
		ymajorgrids=true,
		grid style=dashed,
		scaled ticks=false,
		legend pos=north east,
		xticklabel style={/pgf/number format/fixed},
		yticklabel style={/pgf/number format/.cd, fixed, fixed zerofill, precision=1, /tikz/.cd},
		xlabel={$\text{Area}_{con}$ Ratio},
		xlabel near ticks,
		legend style={
			draw=none,
			at={(-0.17,1.2)},
			anchor=west,
			legend columns=-1,
		}
		]
		
		\pgfplotstableset{
			create on use/x rel/.style={
				create col/expr={
					\thisrow{0}
				}
			},
			create on use/y rel/.style={
				create col/expr={
					\thisrow{1}
				}
			}
		}

		\addplot +[mygreen, line width=1pt, mark=triangle, mark options={scale=1.0, fill=black},text mark as node=true ] table [x ={a}, y={cpd}] {cpd-HEDALSA2.dat};	
		\addplot +[myred, line width=1pt, mark=square, mark options={scale=0.5, fill=mymiddle},text mark as node=true] table [x ={a}, y={cpd}] {cpdOursA2.dat};
            \addplot +[mypurple, line width=1pt, mark=star, mark options={scale=1.0, fill=mypurple},text mark as node=true ] table [x ={a}, y={cpd}] {cpdGWO-A2.dat};
  \legend{HEDALS,Ours}

	\end{axis}

\end{tikzpicture} \label{fig:AreaconNMED}}	
\caption{Average critical path delay ratios $\text{Ratio}_{cpd}$ generated by our framework, HEDALS \cite{meng2023hedals} and traditional GWO under different area constraints ($\text{Ratio} \times \text{Area}_{con}$).}
\label{underAreaconstraint}
\vspace{-0.015in}
\end{figure}

We further compare the average $\text{Ratio}_{cpd}$ achieved by our work with HEDALS\cite{meng2023hedals} and traditional GWO under 5 different $ER$ constraints ($1\%$, $2\%$, $3\%$, $4\%$, $5\%$)  and 5 different $NMED$ constraints ($0.48\%$, $0.98\%$, $1.47\%$, $1.96\%$, $2.44\%$). 
According to results in \Cref{underconstraint}, as the $ER$ or $NMED$ constraint tightens, our framework consistently achieves greater critical path delay reductions than others.
\Cref{underAreaconstraint} illustrates how the average $\text{Ratio}_{cpd}$ varies with different area constraints ($0.8\times \sim1.2 \times$ $\text{Area}_{con}$) under the loosest $ER$ and $NMED$ constraints.  
The results indicate that our framework outperforms other works in timing optimization across all area constraints.
These achievements demonstrate that our framework can generate approximate circuits with superior performance while meeting diverse accuracy and area requirements.

In summary, by leveraging carefully designed approximate actions and the powerful search capabilities of DCGWO, our framework can better exploit the timing improvement inherent in critical path shortening and the enhancement of gate drive strength.
Additionally, compared to traditional GWO, using the double-chase strategy to further formulate the optimization gradients indeed helps the optimizer find better solutions.
Benefiting from the fast implementation of LACs and the inherent parallelism of GWO, our framework maintains low time consumption despite using PrimeTime \cite{primetime} for accurate timing analysis.

\section{Conclusion}
\label{sec:conclu}
In this work, we propose a timing-driven approximate logic synthesis framework based on DCGWO to effectively optimize circuit timing under $ER$ or $NMED$ constraints.
Its main idea involves using DCGWO to optimize both critical path depth and area to achieve precise and efficient optimal approximate circuit generation and utilizing post-optimization under area constraints to convert area reduction into further timing improvement.
According to the experimental results on open-source designs, under the same error and area constraints, our framework can achieve more critical path delay reduction than existing methods within an acceptable time consumption.

\clearpage
\balance
{
\bibliographystyle{IEEEtran}
\bibliography{Top.bib,aging2.bib} 
}

\end{document}